# Channeling-in channeling-out revisited: selected area electron channeling and electron backscatter diffraction

T. Ben Britton, M. Haroon Qaiser, Ruth M. Birch,

Department of Materials Engineering, University of British Columbia, Vancouver, Canada

ben.britton@ubc.ca

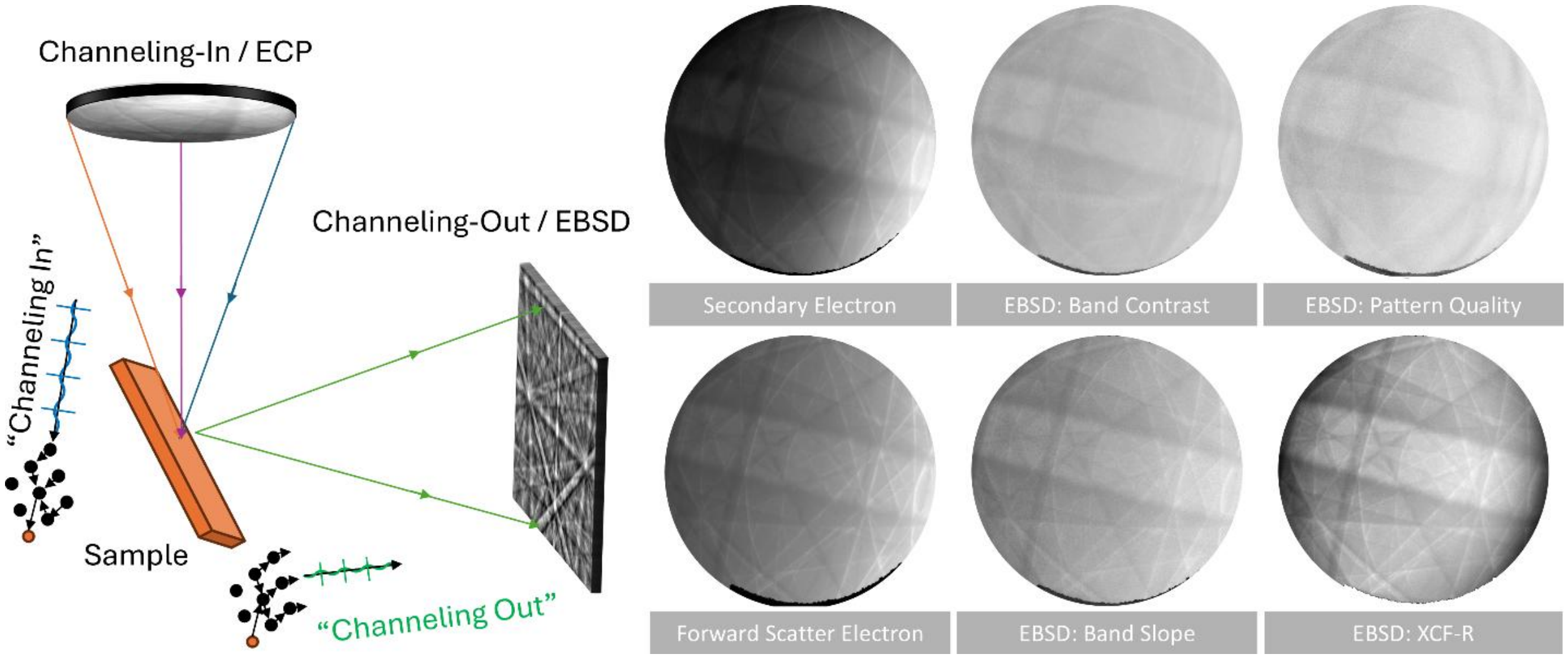

## Abstract

Scanning electron microscopy combined with electron backscatter diffraction (EBSD) and electron channeling provides rich crystallographic contrast, but the mutual influence of channeling-in and channeling-out is often simplified or neglected in quantitative analyses. In this work, we collect selected-area electron channeling patterns (SA-ECPs) acquired from a single-crystal silicon wafer while concurrently recording an EBSD pattern at every incident beam direction, thereby directly probing how channeling-in affects the EBSD signal. We show that common Hough-based EBSD quality metrics (pattern quality, band contrast, and band slope), pattern-matching cross-correlation coefficients, and Fourier-based signal-to-noise ratios all exhibit strong crystallographic modulations that follow the underlying ECP, in both raw and background-corrected patterns. Variations can be up to ~20% of the measured signal. Similar wide-angle channeling features are also visible in conventionally-collect low-magnification EBSD maps, indicating that channeling-in effects are relevant under routine mapping conditions and not only in specialized ECP experiments. These observations highlight that channeling-in can significantly bias quality-based interpretation of EBSD data, with consequences for methods such as pattern blurring analysis, high-resolution strain mapping, and emerging statistical or machine-learning approaches that rely on subtle variations in diffraction patterns. The combined SA-ECP and EBSD strategy presented here offers a practical framework to visualize and potentially control channeling-in/channeling-out coupling in the SEM, suggesting new routes to design experiments and detector configurations that either mitigate or intentionally exploit these dynamical effects.

## Highlights

- Channeling-in strongly modulates EBSD quality metrics.
- Selected-area ECP reveals dependence of EBSD signal to noise ratio.
- Wide-area EBSD maps contain hidden channeling patterns.

## Introduction

A scanning electron microscope (SEM) is a powerful tool that provides detailed information about materials, helping us develop creative solutions to many of today's challenges, such as mineral analysis, metallurgy, ceramics, coatings, and devices. In an SEM, a focused electron beam is scanned across a sample, and signals generated from this interaction are used to form a micrograph. By studying how the electron beam scatters and how different signals are produced, we gain a better understanding of the several types of contrast present in these complex datasets.

SEMs are equipped with an electron gun, which is used to generate a flux of high energy electrons, which is then focussed via a series of lenses. Samples are mapped by deflecting this electron beam (using scan coils) across them. The sample can be moved around using the microscope stage within the SEM vacuum chamber. To collect data about the electron-matter interactions, the scattered electrons are collected using a series of detectors that can be designed to measure secondary electrons (via the Everhart-Thornley Detector or ETD), backscattered electrons, and even pixelated detectors that can be used to collect diffraction patterns for electron backscatter diffraction (EBSD) experiments.

For crystalline materials, the crystal structure (and sample microstructure) can be explored using these methods and the scattering of electrons in crystalline materials has been explored to generate rich simulations that reproduce the intensity variations found within electron channeling patterns (ECPs) [1], electron channeling contrast imaging (ECCI) micrographs [2,3], and EBSD patterns [4,5].

In the present work we discuss 'channeling-in' as the interaction between the primary electron beam and the crystal lattice, which can be strongly affected by, for example: the primary beam voltage; coherence of the electron beam; the convergence angle of the primary beam; and the crystal lattice it interacts with.

In an ECP, typically the beam is when systematically scanned, there is a change the incident beam angle, and vector, with respect to the sample. This means that scanning the beam systematically enables the formation of a ECP-based Kikuchi-pattern. A selected-area (SA-)ECP can be formed when a nearly parallel electron beam is scanned around a common point or area, e.g. via the deflection of the beam through the scan coils and use of the final objective lens to 'rock' the beam around a common point or area [6,7]. A 'wide angle' (WA-)ECP can also be observed for a single crystal sample when the beam is sufficiently parallel and scanned at low magnification [8].

For an ECP, the signal at each point in the captured pattern is formed from a quasi-elastic scattering (from electron-nucleus based phonon interactions). In brief, electrons are emitted from a coherent source, and these electrons can be considered as a series of Bloch-waves. These waves interact with a periodic crystal lattice and can result in Bragg diffraction, and the diffraction is dependant on angle of the incident electron beam (i.e. the plane wave front) together with the structure and crystal orientation of the crystal lattice. Joy [9] provides a comprehensive description of the variation, or modulation, of backscattering signal with regards to the variation in incident angle with respect to a crystal lattice. The ECP is formed from a near surface region and the 'useful' part of the ECP (i.e. the sharp Kikuchi patterns) are near the primary incident electron beam energy [10], and furthermore as the pattern is formed from low loss backscatter electrons it follows that the ECP is formed from the near surface region of a bulk sample.

In practice, angular based contrast within the ECP is dependent on the convergence of the incident electron probe (i.e. a lower convergence angle provides a sharper ECP signal) [9] and the 'quality' of the crystal (i.e. the atomic scattering factors, the Debye-Waller factor of the crystal and the crystal temperature, and any strain gradients within the interaction volume).

Further to ECP formation, electron scattering can also impact channeling-out of the electrons, which produces diffraction patterns which are commonly observed via pixelated detectors for EBSD experiments. In a simplified approximation of EBSD-based diffraction, scattered electrons within the sample now each act as local sources and these sources can interact and cause Kikuchi diffraction [11,12]. While the position of the detector, and type of sample (e.g. thin or thick samples) can vary, similar EBSD-like patterns have now been obtained from a variety of geometries [13].

In short, the relationship between ECP and EBSD contrast is well understood, consisting of a channeling-in, recoil, and channeling-out process. To simplify EBSD pattern analysis, dynamical simulations rely on a model which is the same as used for ECP formation [1], recognizing that modulation of channeling-in can be considered the same as the processes that results in channeling-out contrast for an EBSD pattern [14], via reciprocity.

For both ECP and EBSD pattern simulation, the model described simplifies the approach as for example, the initial inelastic scattering is described as isotropic, located at the atomic positions and scales as $Z^2$, where Z is the atomic number. Next, the high quality diffraction and channeling patterns are simulated using a quasi-elastic scattering many beam diffraction approach where the wave function inside the crystal is taken to be a superposition of (independent) Bloch waves interacting with the unit cell, and adapting convergent beam electron diffraction (CBED) approaches of Zuo and Spence [15]. This quasi-elastic scattering process determines the energy, and thus wavelength, of the Bloch-waves.

Extending this further requires an exploration of the source term for the scattering, recoil, and diffraction events. Vos and Winkleman have performed electrostatic analyser based measurements which indicate that the Kikuchi-like contrast within ECP and EBSD patterns are (primarily) formed from a quasielastic scattering process where the electron loss is dominated by (multiple) electron-phonon interactions [16].

Recognizing the practical nature of these techniques, and the generation of ECPs within a scanning electron microscope where a convergent beam is scanned across the surface of a sample, it is important to also introduce the fact that there is an 'interaction volume.' For SEM-based microanalysis, this volume can be described in multiple ways, but a useful approach is to consider the physical interaction volume which results in contrast that is useful for the technique in question. In the EBSD-community, there is some discussion [17,18] about the particulars of the scattering and diffraction events that result in Kikuchi-like features, but there is agreement that the signal is controlled by Z-number, sample tilt, and primary accelerating voltage [14]. If we consider only the useful electron scattering that produces Kikuchi bands, then the volume of material which results in high-energy backscattered electrons, close to the primary beam energy, is what links EBSD patterns to specific microstructural features. Lower energy regions, which also often result in less coherent scattering, contribute mostly to the diffuse background in the EBSD pattern (which is often subtracted, or less important, for most analysis).

For ECP and ECCI analysis, other scattering contributions can result in changes in contrast, such as Z-contrast, but these may not result in significant crystallographic contrast required for grain and defect identification and imaging. This crystallographic contrast is likely also from low loss backscatter electrons, as recent pattern-matching approaches have verified that high contrast ECPs for flat samples tend to match well when the dynamical simulation is very close to the energy of the primary electron beam [10].

In typical experiments and analysis workflows, the channeling-in and channeling-out phenomena are considered as independent processes, depending on the experiment and analysis approach (e.g. an ECP guided ECCI experiment or regular EBSD analysis). However, Winkelmann et al. [19] conducted an interesting and important experiment with a defocussed beam to produce a plane wave within a LEO 1530 VP field emission gun SEM, where at each point within the scanned map of a sample surface EBSD patterns were collected. This experiment revealed that channeling-in can have a significant impact on the EBSD signal, for example using virtual 'diodes' collected by summing areas of the EBSD pattern and centre of mass (COM) analysis of the brightness distribution of the full (i.e. not background corrected) EBSD pattern. This important result by Winkelmann et al. demonstrates that the interaction volume of an EBSD pattern is influenced by scattering and diffraction effects for the interaction of the incident beam (i.e. channeling-in affects channeling-out).

In the present work, we revisit the impact of channeling-in on channeling-out, but instead of using a defocussed beam, we use regular focussed (and slightly convergent) EBSD beam conditions to collect an EBSD map while the instrument is configured for the formation of an ECP. Our work is motivated by the rise of higher quality detectors and an increased appetite within the community to advance pattern analysis, e.g. using new software algorithms [20,21], and even machine learning methods [22].

## Method

Microscopy was conducted using a TESCAN AMBER-X plasma focussed ion beam scanning electron microscope, equipped with the electron 'channeling' module. This instrument includes a field emission electron source with optics, and scan generation control, which facilitates the collection of electron channeling patterns, using 'channeling mode'. In this mode, both the microscope scan generator and an external third-party scan generator can use the scanning coils to deflect the beam through the final objective lens such that each point within the resulting micrograph has a different incident beam direction on the same selected area (see Figure 1). When the electron beam has a low convergence angle, a sharp ECP is subsequently formed.

Electron backscatter diffraction was conducted using an Oxford Instruments Symmetry S2 detector which is also equipped with diodes for backscatter and forward scattering imaging (upper and lower FSE, respectively), which are mounted around the phosphor screen. Importantly, the Oxford Instruments system is controlled by an external scan generator that can be used while the TESCAN AMBER-X is in channeling mode. In addition to regular EBSD-mapping, this means that EBSD patterns can be concurrently collected for every scanned point used to generate a SA-ECP.

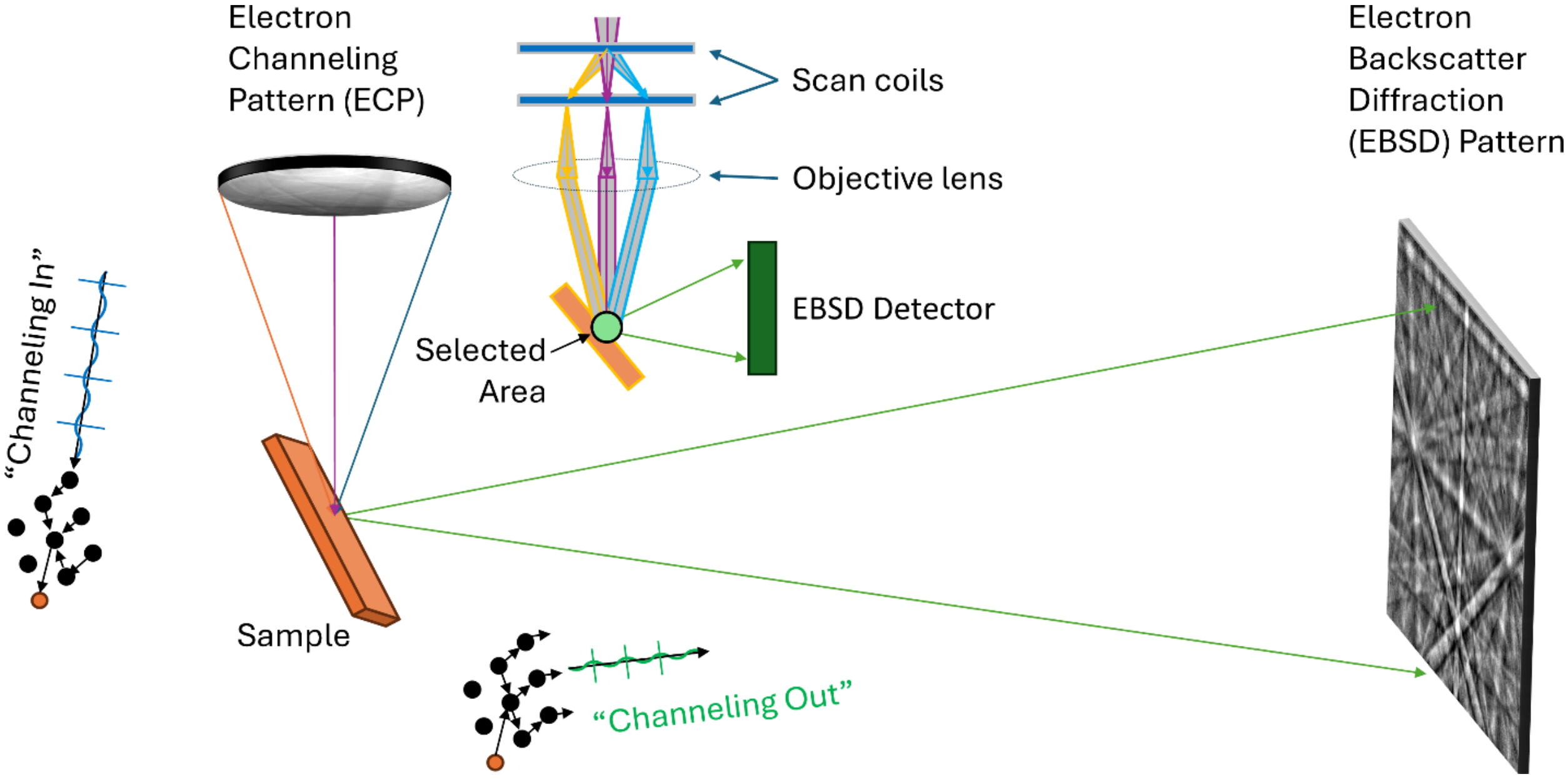

*Figure 1: Schematic of channeling-in and channeling-out when collecting a map using 'channeling' mode on the TESCAN AMBER-X. The electron channeling mode uses the scan coils to deflect the beam through the objective lens such that each point within the ECP-micrograph is related to a systematic variation in the incident beam angle with regards to the sample. Concurrently to ECP collection via a traditional detector, EBSD patterns can be captured at each point*

To perform these experiments, a single crystal of semiconductor grade silicon was mounted on an aluminium stub for analysis. The sample was mounted at 70° on a pre-tilted holder for ECCI and EBSD analysis. This sample is strain free, polished flat and contains very few defects and therefore the only contrast within a subsequent map can be related back to the control of the instrument. In this microscope, the EBSD detector is placed to provide a 'static' configuration to aid 3D EBSD experiments [23], and so the detector is placed at an angle of 14.7° with regards to the primary tilt axis of the SEM.

All electron imaging and analysis was conducted at a primary beam energy of 20 keV. EBSD and ECP experiments were conducted with a 1 nA probe current, which results in a half beam convergence angle of 3.37 mrad (as calculated using a model that is included within the Python API of the TESCAN AMBER-X instrument). For EBSD pattern capture, the Symmetry S2 detector was operated in Resolution Mode (1244x1024 pixel patterns), and this enables high bit depth images to be collected with a high angular resolution. Patterns were captured at a pattern collection rate of 38.25 Hz. This results in a total incident electron dose on the sample of $1.63 \times 10^{8}$ electrons per point. Forward scatter electron (FSE) micrographs were also captured from the five diodes located around the EBSD detector.

Regular SEM-based imaging was used to collect an ECP using the TESCAN scan generator, while the microscope was in channeling mode.

Subsequently, two maps were collected:

- ECP-EBSD map captured while the microscope was switched to channeling mode. The EBSD mapping software, Aztec, asked the microscope to move the beam from position to position within the frame that was observed on the SEM computer. This frame shows an electron channelling pattern, and each mapped X-Y position captured the signal of the electron striking the sample at a different angle (and as we shall see, position). Note that the dynamic focus correction, scan rotation, and tilt correction features were disabled.
- Large area EBSD map, using the conventional EBSD-based mapping mode. The SEM was scanned across the surface of the sample, and the microscope was configured such that scanning grid was rotated to present a 'regular' EBSD mapping geometry (i.e. the map horizontal axis is aligned with the sample tilt axis).

For both experiments, EBSD patterns were all saved to disk and EBSD data was saved to the h5oina format [24]. Both 'unprocessed' and 'processed' patterns were saved to disk. The unprocessed data are the raw patterns that are collected by the detector, and the processed data includes background correction (which is performed using standard routines within the Oxford Instruments software, including a hardware collection of the background using a copper test artifact). Online Hough/Radon based pattern analysis was performed using Aztec, and pattern matching was performed within MapSweeper using Aztec Crystal, and subsequently imported into MATLAB for analysis using MTEX based scripts [25].

To aid indexing of the ECP, electron channeling pattern simulation of Si was performed using EMSoft v5.0 [26] and a 5000 × 5000 pixel Kikuchi stereogram was generated for 20 keV primary beam energy, keeping lower threshold of $d_{hkl}$ as 0.2 Å to include more plane reflectors using a dynamical simulation. Subsequent ECP reprojection, indexing and band analysis was performed using the open-source graphical user interface AstroECP [10] in MATLAB R2023b.

## Results

SA-ECP analysis was performed to verify that the TESCAN and Oxford Instruments (OI) scan generators collected similar data, as shown in Figure 2, so that this data could be correctly interpreted. Using AstroECP the patterns were indexed successfully, and the pole-figures and unit cell are consistent with the orientation of the sample within the electron microscope chamber. All crystallographic features measured are correctly accounted for, noting that the SEM scan directions (x' and z') are rotated 15.5° rotated about the optic axis of the microscope, and this is consistent with a map collected in the absence of scan rotation and positioned for EBSD analysis using the detector configuration on this microscope.

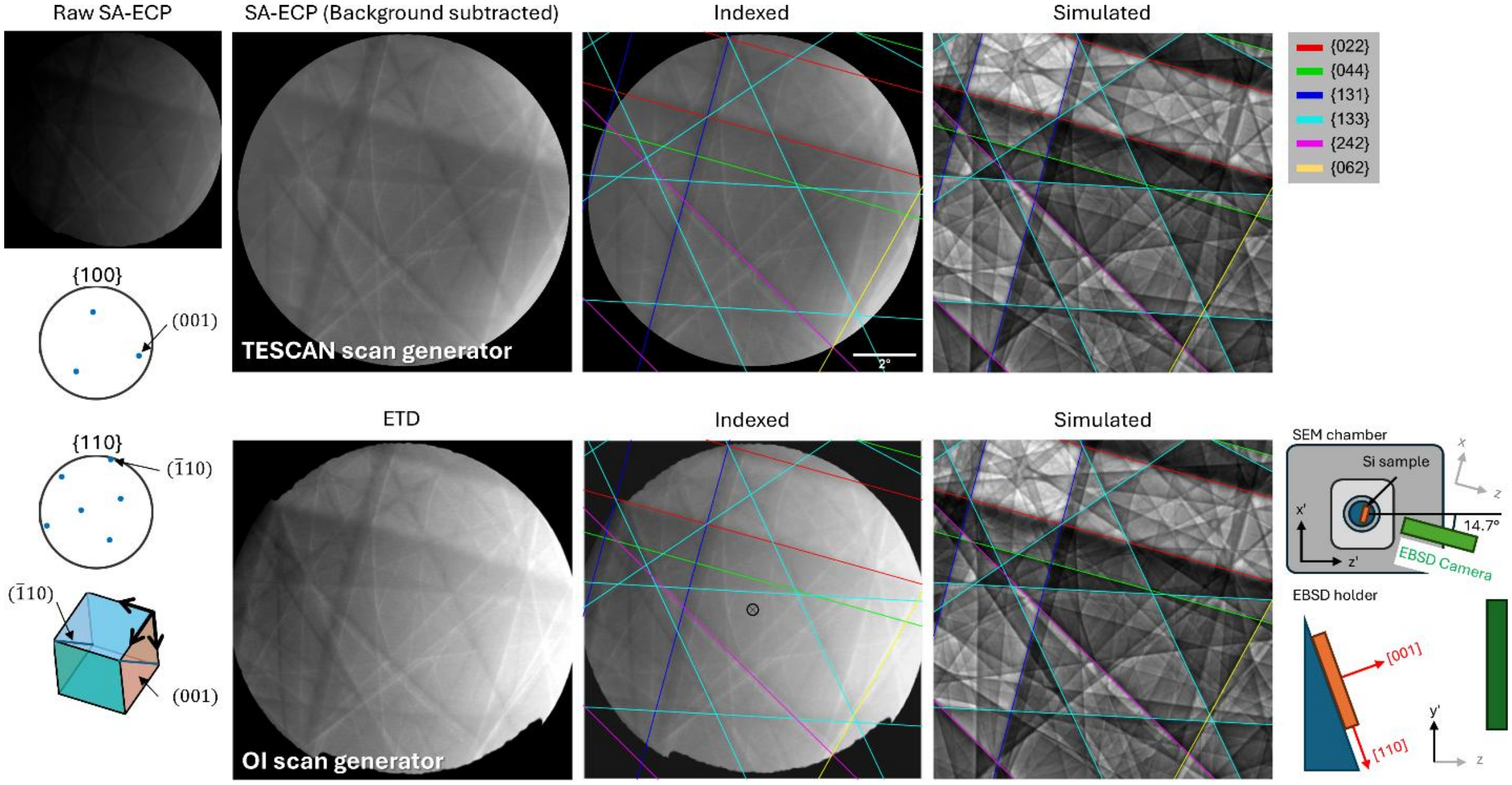

*Figure 2: Collection of the selected area electron channeling patterns (SA-ECP) using the TESCAN Essence Scan Generator and the Oxford Instruments external scan generator, while the AMBER-X is placed in 'channeling' mode. Patterns were indexed using AstroECP and compared to re-projected simulated patterns which were created using dynamical diffraction theory. The relationship between each pattern and the experimental chamber is also shown.*

When the OI external scan generator was used and the ETD signal was collected, concurrently the five 'backscatter' and 'forward' scatter diodes mounted around the EBSD detector collected a signal (Figure 3). These signals again show strong crystallographic contrast, especially on the lower three diodes and the ECPs formed are similar in structure to the ETD detector micrograph. The 'upper' pair of micrographs show less ECP contrast than the 'lower' three diodes, which is consistent with upper diodes usually showing less crystallographic contrast for this detector insertion position (similar to findings in [27]).

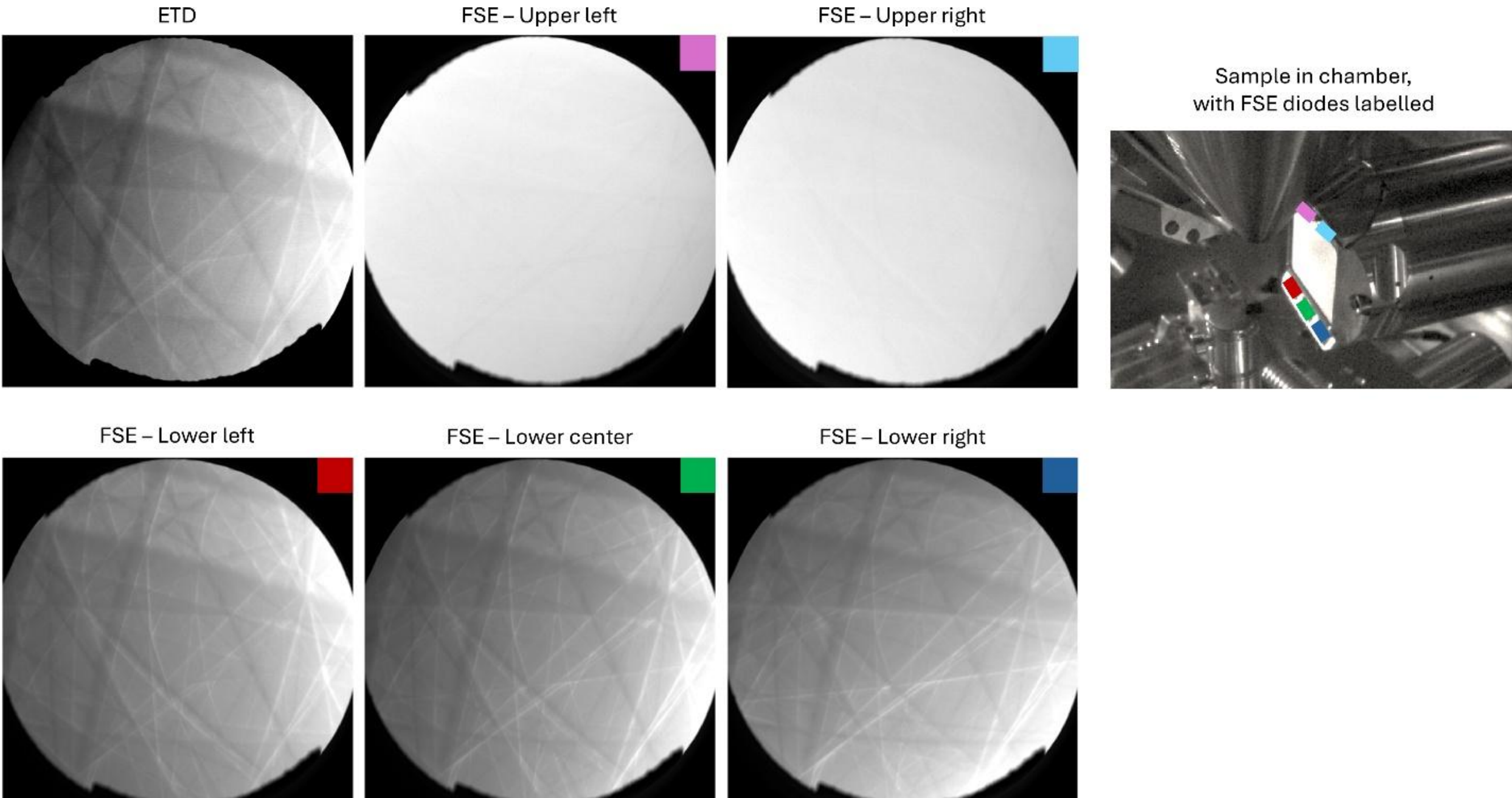

*Figure 3: A comparison of the ETD detector micrograph and the five diodes mounted around the EBSD detector, as collected using the Oxford Instruments external scan generator while the micrograph was in channeling mode. All micrographs show strong crystallographic features, revealing the influence of electron channeling in on these signals.*

Next, the quality metrics for the EBSD pattern map collected were assessed (Figure 4). Remarkably, the band contrast, band slope, pattern quality, and cross-correlation metrics showed significant crystallographic contrast associated with features that are also found within the ETD and FSE diode-based ECPs. While the commercial-basis of the normalization of these metrics is unclear, the distribution of values within the ECP are significant: the band contrast varies between ~180 and ~240 (i.e. mean +/- 15% signal range); the band slope varies between ~150 and ~240 (i.e. mean +/- 12.5%); pattern quality varies between ~640 and 820 (i.e. mean +/- 6%); and the cross-correlation values varies between ~0.35 and ~0.55 (i.e. mean +/- 22%).

The mean angular deviation and kernel average misorientation maps show slight variations at the extremes of the ECP, and this is related to the size of the selected area pattern and how the pattern centre model is incorrect for the indexing of the points within this map. Importantly, these maps do not contain higher spatial frequency features that are seen within the other ECPs.

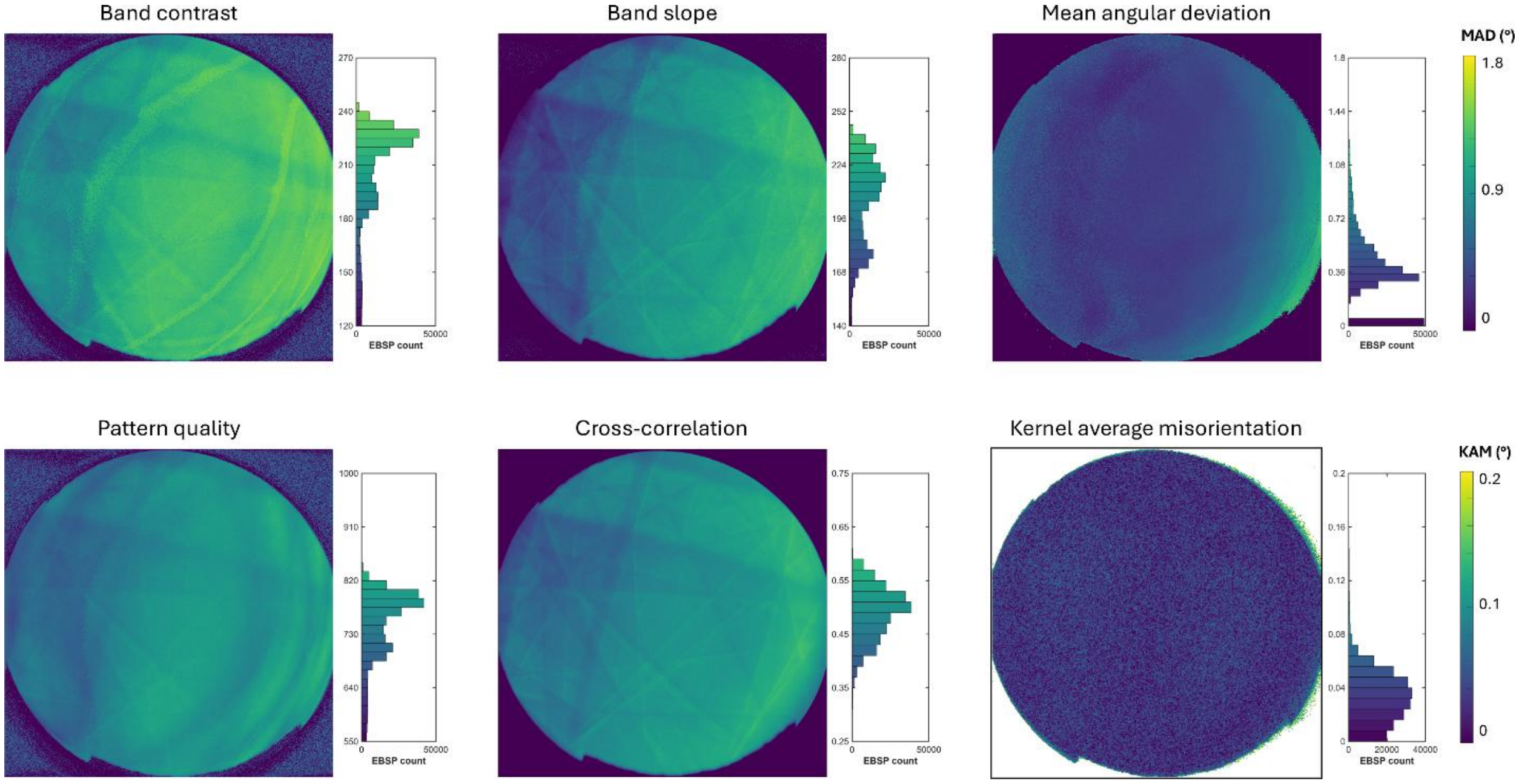

*Figure 4: Analysis of EBSD pattern metrics for an EBSD map collected while the microscope is in channeling mode. The band contrast, band slope, pattern quality and (Pattern Matching) cross-correlation metrics all show strong ECP-related features. The Hough-based quality metric – mean angular deviation (MAD) shows limited correlation with the ECP and instead shows a slightly variation due to the poor-quality dynamic pattern centre model. The Pattern Matching based Kernel average misorientation data shows no ECP-related contrast.*

Next, as shown in Figure 5, analysis of individual EBSD patterns was performed for regions within the ECP which are close together but show a variation in contrast within the ECP (e.g. either side of the Kikuchi band edge). Here the pattern pairs reveal that there is a slight variation in contrast typically towards the bottom of the EBSD pattern, and that the pattens from the 'bright' ECP regions typically contain slightly higher frequency information within the Fourier power spectra. For a sample which is tilted to a high angle, such as here, the bottom of the EBSD pattern likely contains information from electrons that have travelled further in the sample (and lost more energy) [17].

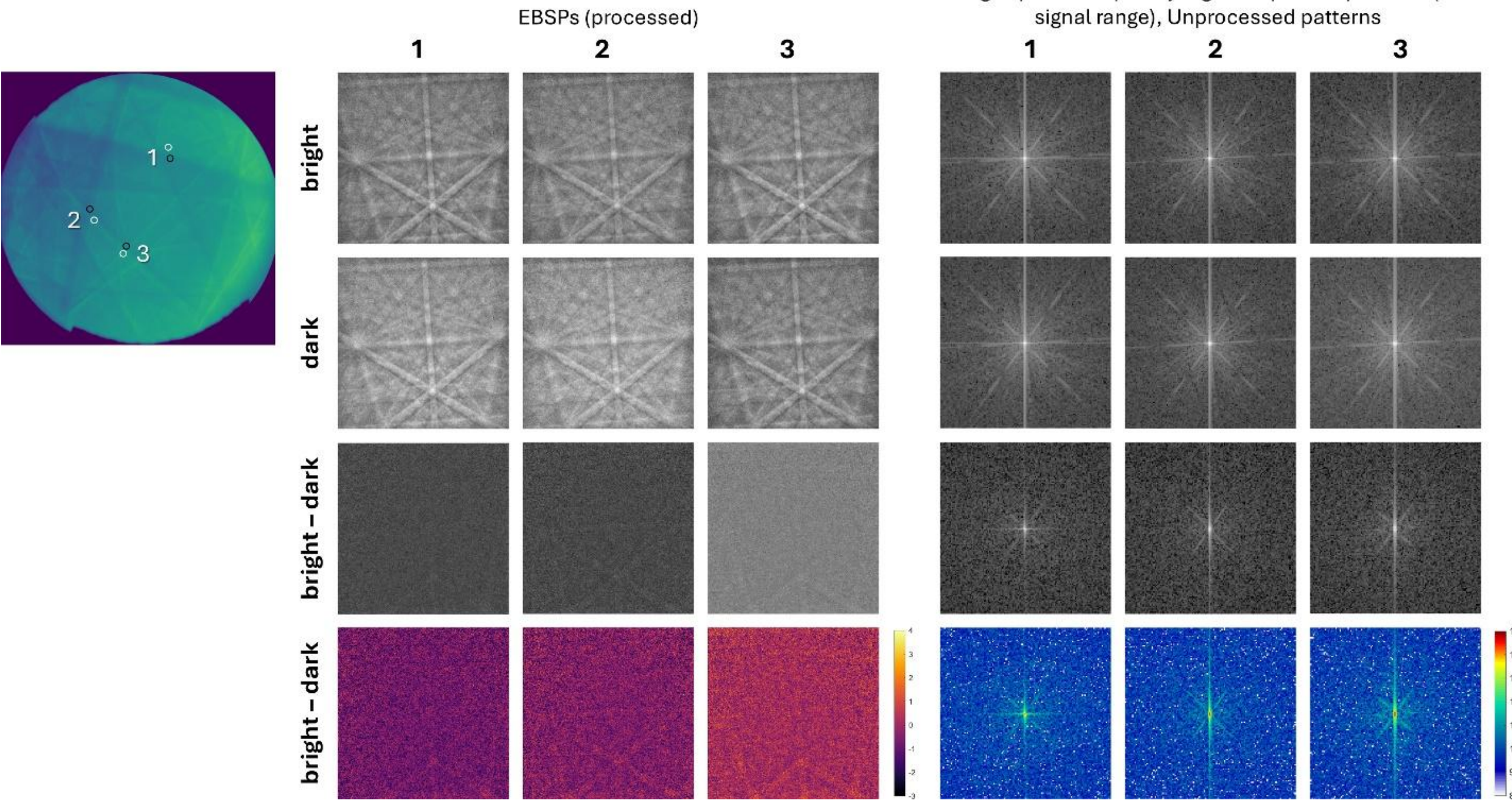

*Figure 5: Analysis of EBSD pattern contrast at comparison points in the SA-ECP map shows that the diffraction patterns change in contrast as the electron beam is channeled-in differently for either side of major Kikuchi band edges.*

Analysis of these patterns motivates further analysis of the signal to noise within the EBSD patterns, using Fourier power spectrum analysis (Figure 6). In this analysis, the information within the 'signal' region of the FFT-power spectrum is radially averaged and measured as a fraction of the radial average of the noise spectrum. The signal to noise maps of both the unprocessed and processed EBSD patterns were analysed here, and both maps reveal the ECP clearly. The unprocessed data shows significant topographical effects (as there is a long-range gradient which extends 'down' the physical sample in the ECP).

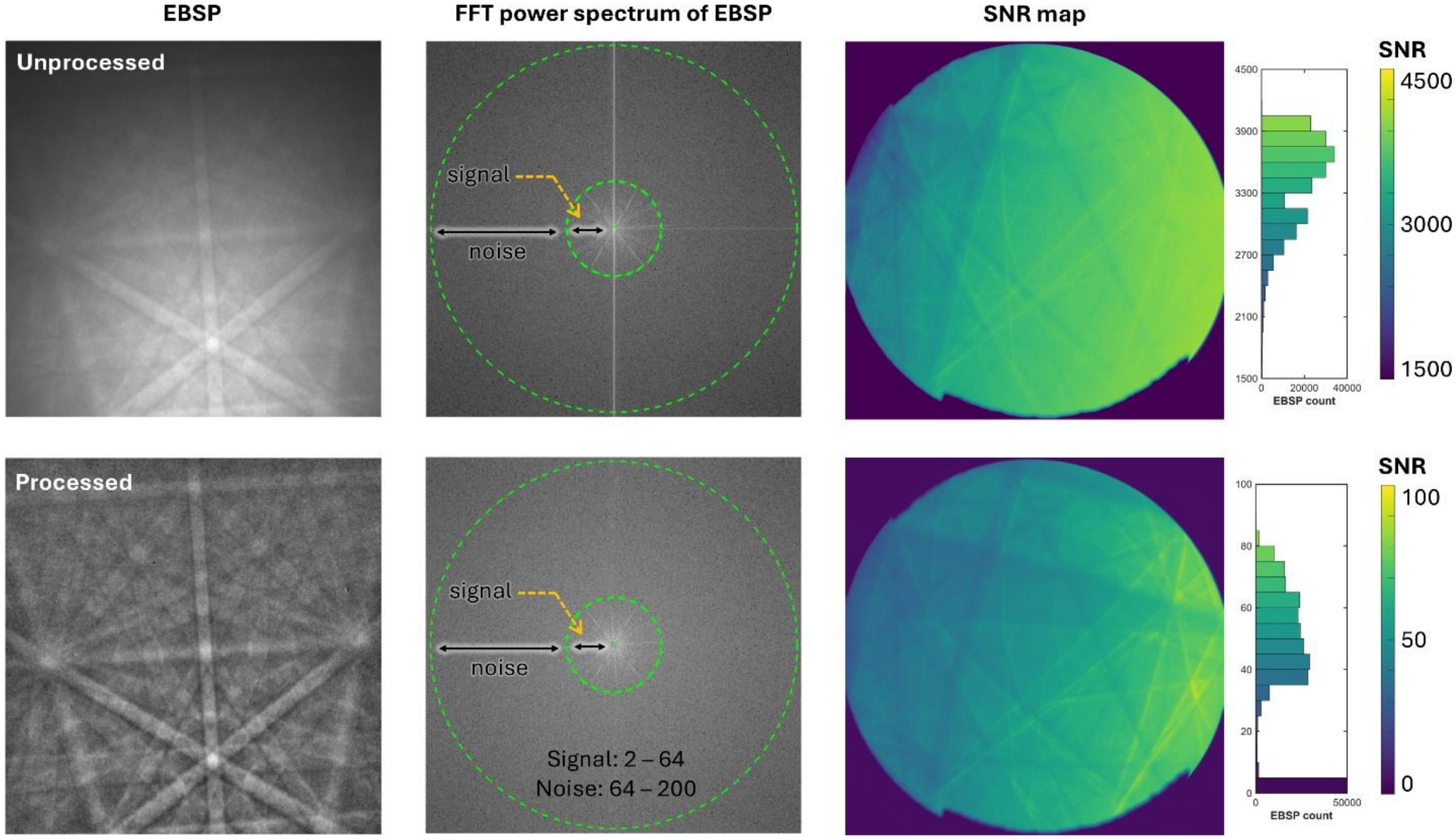

*Figure 6: Fourier based signal to noise analysis of the unprocessed (i.e. 'raw') and processed (i.e. background corrected) EBSD patterns collected show that the ECP is present with both patterns.*

The large area silicon EBSD map was analysed in an equivalent way. As a reminder, this map was collected using a more regular EBSD-set up (i.e. channeling mode was switched off) and the beam was scanned at low magnification. Figure 7 reveals that there is crystallographic contrast within this map, which is related to the 'wide angle' electron channeling patterns that form due to scanning the electron beam across a large area. For example, the {400} band can be seen running vertically down the resulting micrograph in Figure 7. Indexing of this pattern with more confidence is complicated, both as the pattern subtends a narrow capture angle and the sample was tilted (which means that the pattern is stretched as compared to a wide-angle selected area channeling pattern captured at 0° sample tilt).

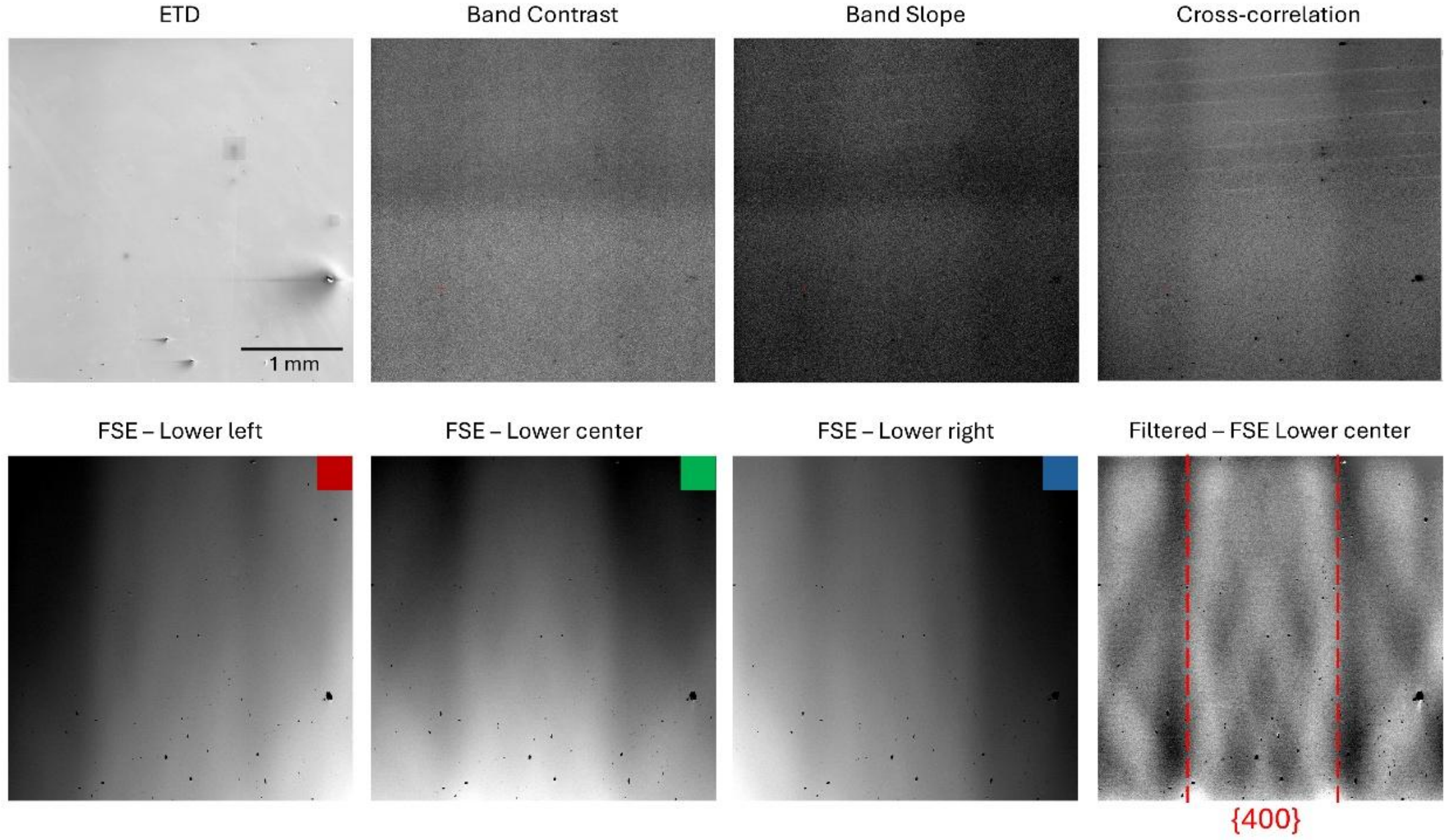

*Figure 7: Example large area silicon EBSD map, which reveals the presence of a wide angle ECP within the underlying forward scatter electron (FSE) micrographs and the EBSD quality metrics (band contrast, band slope, and cross-correlation). After Fourier-based filtering the characteristic features of the electron channeling-in pattern can be observed, including the vertical {400} band.*

## Discussion

The present work explores the variation in signals formed during an EBSD-like experiment, where the electron beam has been systematically varied over a large angular range through the formation of an ECP. The key finding is that there is a significant impact of channeling-in on the EBSD-pattern analysis, including the variation in Hough-based pattern quality metrics (such as pattern quality, band contrast and band slope), in pattern matching quality from cross correlation with simulated patterns, and through direct signal to noise analysis of both the processed and unprocessed patterns. These variations are also seen within large area maps taken from the same single crystal of silicon, and the wide area electron channeling pattern is observed.

The results indicate that channeling-in can play a significant role in the analysis of EBSD patterns, including basic metrics as well as more subtle variations within the patterns. These dynamical effects are similar, in part, to the effects seen in a transmission electron microscope (TEM) and the rise of precession scanning generators to simplify the analysis of TEM-based diffraction patterns [28]. In the TEM, an additional scan generator system can be used to alter the incident angle of the electron beam, scanning in a small precession circle. In practice the precession circle diameter is chosen to balance the spatial resolution of the illuminated volume in the thin foil while also providing a sufficient range of incident beam angles that dynamical effects are cancelled out. In the SEM, in theory a scan generation system could be used in a comparable manner to suppress the additional complexity that channeling-in imposes for EBSD analysis. For example, summing a small region of the ECP experiment can provide a range of incident angles for the EBSD pattern. However, objective lens aberrations make it difficult to maintain an ECP-like precession area on the sample, especially since this area becomes even larger under the high tilts used in typical EBSD experiments, but this could be improved with either improved electron-optics control and online software correction [29].

Moreover, due to the reciprocal nature of electron scattering, channeling-out effects can also influence channeling-in during ECP generation and interpretation, as well as for electron channeling contrast imaging (ECCI). In these cases, the effects are likely to be small as the angle subtended by each forescatter diode (or the BSE diode segments for more conventional ECP analysis) is large. However, when pixelated sensors are used to generate virtual FSE images [19,27,30,31], especially as more complicated masks are explored [32], this assumption may no longer be true and consideration of channelling-in and channeling-out may occasionally be required. In the limit, use of machine learning methods which include the development of latent space based analysis of EBSD patterns are also a version of adaptative masking, and again channeling-in and channeling-out effects likely impact the data [22].

Another area where channeling-in affects EBSD patterns analysis concerns efforts to quantify pattern blurring, such as dislocation density calculations which aimed to adapt X-ray diffraction line profile methods [33], but at the electron scale. To recap, early work by Wilkinson and Dingley explored calibrated analysis of pattern blurring by FFT power spectrum analysis of EBSD pattern features [34]. Recently, there has been a re-interest in this approach as detectors improve in quality and there is increased access to compute resource for large scale diffraction pattern analysis and the rise in quality of pattern simulation tools [35,36].

In the present work, an indirect detector has been used which consists of a polycrystalline phosphor screen to enable collection of the resulting diffraction patterns. There has been a recent growth in direct electron detector-based EBSD analysis, where the detector can be made from a single crystal of silicon. These detectors include detector channeling effects where a 'negative' channeling pattern can be observed [37,38] depending on the relative strengths of input signal, and this can be a benefit if used for calibration of the system (i.e. the detector pattern provides direct access to the orientation of the diffraction camera and the radiative point source [39,40]). In practical terms, this signal may be a 'third order' modulation of the resulting signal, but efforts to use DED's for more advanced analysis are only emerging. If the detector pattern is significant, given the EBSD communities experience in adapting pixelated detectors, there is also potential to use an amorphous silicon detector [41].

Successive channeling events add complexity often overlooked in past studies, because both instrumentation and analysis were simpler (and typically sufficient). Yet recognizing where these effects occur allows us to design experiments that either mitigate or leverage them. Here we can draw some inspiration from the TEM community, both in terms of the use of precession electron diffraction (as previously discussed), and also the use of structured apertures to control the illumination and optimize the analysis (e.g. for strain mapping in 4D-STEM [42] and the use of electron vortex beams [43]). In these cases, the shape of the probe is increasingly important to control the quality of the experiment, and to optimize the desired contrast mechanism.

Notably in this work, we do not explore the influence of this on polycrystalline samples as controlling those experiments to generate additional insight is difficult, as compared to the silicon case study here. The reciprocal nature of channeling-in and channeling-out has been discussed briefly for polycrystals, using EBSD-informed descriptions of ECCI-features found in polycrystalline metals [31]. However, these effects will still be present and should be considered when making claims or developing new analysis methods.

## Conclusion

The major conclusions of the present work are that channeling-in conditions systematically modulate EBSD pattern quality metrics, including band contrast, band slope, pattern quality, and pattern-matching cross-correlation. Depending on the probe shape, scanning strategy and sample variations could be up to ~20% of the measured signal. Selected-area electron channeling pattern experiments, where EBSD patterns were collected for every incident beam direction, directly visualize how angular variations in channeling-in affect both conventional Hough-based metrics and Fourier-based signal-to-noise measures in raw and background-corrected patterns.

Similar ECP-related modulations are also evident in low-magnification, large-area EBSD maps through wide-angle channeling patterns, indicating that channeling-in effects may influence routine mapping conditions more significantly than often recognized.

These observations suggest that advanced quantitative EBSD analyses - such as those involving pattern blurring for defect density estimation, high-resolution strain mapping, or data-driven and machine-learning approaches - should explicitly account for channeling-in as a potential confounding or exploitable contrast mechanism.

Finally, the combined experimental strategy of selected-area ECPs with concurrent EBSD acquisition offers a pathway to probe channeling-in and channeling-out effects, and, with refinement of the selected-area approach, could present a means to suppress channeling-in influences on EBSD measurements.

## Acknowledgements

The authors thank Jiří Dluhoš and Václav Ondračka, both from TESCAN, for helpful discussions regarding the scan generation system on the TESCAN AMBER-X microscope. The authors would like to thank a range of funders that supported this collaborative work: Natural Sciences and Engineering Research Council of Canada (NSERC) [Discovery grant: RGPIN-2022-04762, 'Advances in Data Driven Quantitative Materials Characterization'] (TBB and HQ); the Canada Research Chair program ['Advancing Correlative Multimodal Electron Microscopy'] (TBB); PEEF-2023 (HQ). Electron microscopy was performed within the Electron Microscopy Laboratory at the University of British Columbia, supported by funding from British Columbia Knowledge Fund (BCKDF) Canada Foundation for Innovation – Innovation Fund (CFI-IF) [#39798, 'AM+']. This research was supported in part through the computational resources and services provided by Advanced Research Computing (ARC) at University of British Columbia (UBC), Canada.

## CRediT author statement

Ben Britton – Conceptualization, Methodology, Investigation, Resources, Writing – Original Draft, Writing – Reviewing and Editing, Supervision, Funding Acquisition

Haroon Qaiser – Investigation, Writing – Reviewing and Editing, Visualization

Ruth Birch – Investigation, Writing – Reviewing and Editing

## References

[1] A. Winkelmann, B. Schröter, W. Richter, Dynamical simulations of zone axis electron channelling patterns of cubic silicon carbide, Ultramicroscopy 98 (2003) 1–7. https://doi.org/10.1016/S0304-3991(03)00021-4.

[2] E. Pascal, B. Hourahine, G. Naresh-Kumar, K. Mingard, C. Trager-Cowan, Dislocation contrast in electron channelling contrast images as projections of strain-like components, Materials Today: Proceedings 5 (2018) 14652–14661. https://doi.org/10.1016/j.matpr.2018.03.057.

[3] Y.N. Picard, M. Liu, J. Lammatao, R. Kamaladasa, M. De Graef, Theory of dynamical electron channeling contrast images of near-surface crystal defects, Ultramicroscopy 146 (2014) 71–78. https://doi.org/10.1016/j.ultramic.2014.07.006.

[4] P.G. Callahan, M. De Graef, Dynamical Electron Backscatter Diffraction Patterns. Part I: Pattern Simulations, Microanal 19 (2013) 1255–1265. https://doi.org/10.1017/S1431927613001840.

[5] A. Winkelmann, Dynamical Simulation of Electron Backscatter Diffraction Patterns, in: A.J. Schwartz, M. Kumar, B.L. Adams, D.P. Field (Eds.), Electron Backscatter Diffraction in Materials Science, Springer US, Boston, MA, 2009: pp. 21–33. https://doi.org/10.1007/978-0-387-88136-2_2.

[6] C.G. Van Essen, E.M. Schulson, R.H. Donaghay, Electron Channelling Patterns from Small (10 μm) Selected Areas in the Scanning Electron Microscope, Nature 225 (1970) 847–848. https://doi.org/10.1038/225847a0.

[7] C.G. Van Essen, E.M. Schulson, Selected area channelling patterns in the scanning electron microscope, J Mater Sci 4 (1969) 336–339. https://doi.org/10.1007/BF00550403.

[8] D.G. Coates, Kikuchi-like reflection patterns obtained with the scanning electron microscope, The Philosophical Magazine: A Journal of Theoretical Experimental and Applied Physics 16 (1967) 1179–1184. https://doi.org/10.1080/14786436708229968.

[9] Electron Channeling Patterns, in: Techniques in Physics, Academic Press, 1989: pp. 69–118. https://doi.org/10.1016/B978-0-12-353855-0.50009-9.

[10] M.H. Qaiser, L. Berners, R.J. Scales, T. Zhang, M. Heller, J. Dluhoš, S. Korte-Kerzel, T.B. Britton, AstroECP: towards more practical electron channeling contrast imaging, J Appl Cryst 59 (2026). https://doi.org/10.1107/S1600576726000567.

[11] A. Winkelmann, Dynamical effects of anisotropic inelastic scattering in electron backscatter diffraction, Ultramicroscopy 108 (2008) 1546–1550. https://doi.org/10.1016/j.ultramic.2008.05.002.

[12] A. Winkelmann, C. Trager-Cowan, F. Sweeney, A.P. Day, P. Parbrook, Many-beam dynamical simulation of electron backscatter diffraction patterns, Ultramicroscopy 107 (2007) 414–421. https://doi.org/10.1016/j.ultramic.2006.10.006.

[13] T. Zhang, L. Berners, J. Holzer, T.B. Britton, Comparison of Kikuchi diffraction geometries in scanning electron microscope, Materials Characterization (2025) 114853. https://doi.org/10.1016/j.matchar.2025.114853.

[14] L. Reimer, Scanning Electron Microscopy: Physics of Image Formation and Microanalysis, Second Edition, Meas. Sci. Technol. 11 (2000) 1826. https://doi.org/10.1088/0957-0233/11/12/703.

[15] J.M. Zuo, J.C.H. Spence, Electron Microdiffraction, Springer Science & Business Media, 2013.

[16] A. Winkelmann, M. Vos, Site-Specific Recoil Diffraction of Backscattered Electrons in Crystals, Phys. Rev. Lett. 106 (2011) 085503. https://doi.org/10.1103/PhysRevLett.106.085503.

[17] A. Winkelmann, T.B. Britton, G. Nolze, Constraints on the effective electron energy spectrum in backscatter Kikuchi diffraction, Physical Review B 99 (2019) 064115.

[18] N.M. Della Ventura, K. Moore, M.P. Echlin, M.R. Begley, T.M. Pollock, M. De Graef, D.S. Gianola, Energy-resolved EBSD using a monolithic direct electron detector, Ultramicroscopy 281 (2026) 114301. https://doi.org/10.1016/j.ultramic.2025.114301.

[19] A. Winkelmann, G. Nolze, S. Vespucci, G. Naresh-Kumar, C. Trager-Cowan, A. Vilalta-Clemente, A. j. Wilkinson, M. Vos, Diffraction effects and inelastic electron transport in angle-resolved microscopic imaging applications, Journal of Microscopy 267 (2017) 330–346. https://doi.org/10.1111/jmi.12571.

[20] Q. Shi, L. Jiao, D. Loisnard, C. Dan, Z. Chen, H. Wang, S. Roux, Improved EBSD indexation accuracy by considering energy distribution of diffraction patterns, Materials Characterization 188 (2022) 111909. https://doi.org/10.1016/j.matchar.2022.111909.

[21] T. Vermeij, J.P.M. Hoefnagels, A consistent full-field integrated DIC framework for HR-EBSD, Ultramicroscopy 191 (2018) 44–50. https://doi.org/10.1016/j.ultramic.2018.05.001.

[22] M. Calvat, C. Bean, D. Anjaria, H. Park, H. Wang, K. Vecchio, J.C. Stinville, Learning metal microstructural heterogeneity through spatial mapping of diffraction latent space features, Npj Comput Mater 11 (2025) 284. https://doi.org/10.1038/s41524-025-01770-8.

[23] R. Birch, S. Li, S. Sharang, W.J. Poole, B. Britton, Large volume 'chunk' lift out for 3D tomographic analysis using analytical plasma focussed ion beam – scanning electron microscopy, Micron 203 (2026) 103986. https://doi.org/10.1016/j.micron.2025.103986.

[24] oinanoanalysis/h5oina, (2025). https://github.com/oinanoanalysis/h5oina (accessed March 9, 2026).

[25] F. Bachmann, R. Hielscher, H. Schaeben, Texture Analysis with MTEX – Free and Open Source Software Toolbox, Solid State Phenomena 160 (2010) 63–68. https://doi.org/10.4028/www.scientific.net/SSP.160.63.

[26] S. Singh, M. De Graef, Dictionary Indexing of Electron Channeling Patterns, Microanal 23 (2017) 1–10. https://doi.org/10.1017/S1431927616012769.

[27] T.B. Britton, D. Goran, V.S. Tong, Space rocks and optimising scanning electron channelling contrast, Materials Characterization 142 (2018) 422–431.

[28] P.A. Midgley, A.S. Eggeman, Precession electron diffraction – a topical review, IUCrJ 2 (2015) 126–136. https://doi.org/10.1107/S2052252514022283.

[29] J. Guyon, H. Mansour, N. Gey, M.A. Crimp, S. Chalal, N. Maloufi, Sub-micron resolution selected area electron channeling patterns, Ultramicroscopy 149 (2015) 34–44. https://doi.org/10.1016/j.ultramic.2014.11.004.

[30] S.I. Wright, M.M. Nowell, R. de Kloe, P. Camus, T. Rampton, Electron imaging with an EBSD detector, Ultramicroscopy 148 (2015) 132–145. https://doi.org/10.1016/j.ultramic.2014.10.002.

[31] N. Brodusch, H. Demers, R. Gauvin, Imaging with a Commercial Electron Backscatter Diffraction (EBSD) Camera in a Scanning Electron Microscope: A Review, Journal of Imaging 4 (2018) 88. https://doi.org/10.3390/jimaging4070088.

[32] N.M. della Ventura, J.D. Lamb, W.C. Lenthe, M.P. Echlin, J.T. Pürstl, E.S. Trageser, A.M. Quevedo, M.R. Begley, T.M. Pollock, D.S. Gianola, M. De Graef, Orientation-adaptive virtual imaging of defects using EBSD, Ultramicroscopy 276 (2025) 114205. https://doi.org/10.1016/j.ultramic.2025.114205.

[33] B.E. Warren, B.L. Averbach, The Separation of Cold-Work Distortion and Particle Size Broadening in X-Ray Patterns, J. Appl. Phys. 23 (1952) 497. https://doi.org/10.1063/1.1702234.
[34] A.J. Wilkinson, D.J. Dingley, Quantitative deformation studies using electron back scatter patterns, Acta Metallurgica et Materialia 39 (1991) 3047–3055. https://doi.org/10.1016/0956-7151(91)90037-2.
[35] X. Li, X. Li, T. Wu, C. Lv, C. Cai, EBSD patterns simulation of dislocation structures based on electron diffraction dynamic theory, Micron 169 (2023) 103461. https://doi.org/10.1016/j.micron.2023.103461.
[36] C. Zhu, M. De Graef, EBSD pattern simulations for an interaction volume containing lattice defects, Ultramicroscopy 218 (2020) 113088. https://doi.org/10.1016/j.ultramic.2020.113088.
[37] T. Zhang, R.M. Birch, G.J. Francolini, E. Karakurt Uluscu, B. Britton, Practical Considerations for Crystallographic and Microstructure Mapping With Direct Electron Detector-Based Electron Backscatter Diffraction, Microanal 31 (2025) ozaf076. https://doi.org/10.1093/mam/ozaf076.
[38] S. Vespucci, G. Naresh-Kumar, C. Trager-Cowan, K.P. Mingard, D. Maneuski, V. O'Shea, A. Winkelmann, Diffractive triangulation of radiative point sources, Appl. Phys. Lett. 110 (2017) 124103. https://doi.org/10.1063/1.4978858.
[39] S. Vespucci, A. Winkelmann, K. Mingard, D. Maneuski, V. O'Shea, C. Trager-Cowan, Exploring transmission Kikuchi diffraction using a Timepix detector, J. Inst. 12 (2017) C02075. https://doi.org/10.1088/1748-0221/12/02/C02075.
[40] T. Zhang, R. Gauvin, A. Winkelmann, T.B. Britton, Dynamical Simulation of On-axis Transmission Kikuchi and Spot Diffraction Patterns, Based on Accurate Diffraction Geometry Calibration, (2026). https://doi.org/10.48550/arXiv.2603.22628.
[41] J. Chabbal, C. Chaussat, T. Ducourant, L. Fritsch, J. Michailos, V. Spinnler, G. Vieux, M. Arques, G. Hahm, M. Hoheisel, H. Horbaschek, R.F. Schulz, M.F. Spahn, Amorphous silicon x-ray image sensor, in: Medical Imaging 1996: Physics of Medical Imaging, SPIE, 1996: pp. 499–510. https://doi.org/10.1117/12.237812.
[42] C. Mahr, K. Müller-Caspary, T. Grieb, F.F. Krause, M. Schowalter, A. Rosenauer, Accurate measurement of strain at interfaces in 4D-STEM: A comparison of various methods, Ultramicroscopy 221 (2021) 113196. https://doi.org/10.1016/j.ultramic.2020.113196.
[43] L. Clark, A. Béché, G. Guzzinati, A. Lubk, M. Mazilu, R. Van Boxem, J. Verbeeck, Exploiting Lens Aberrations to Create Electron-Vortex Beams, Phys. Rev. Lett. 111 (2013) 064801. https://doi.org/10.1103/PhysRevLett.111.064801.

## Data Statement

Data is available on Zenodo <link added after peer review>.

## Generative AI Statement

After a first draft was written, TBB used Claude Opus 4.6 to improve the grammar and flow of this manuscript. After using this tool/service, the authors reviewed and edited the content as needed and take full responsibility for the content of the published article.